\documentclass[aps,physrev,twocolumn,superscriptaddress]{revtex4-2}
\pdfoutput=1

\usepackage{graphicx}
\usepackage{amsfonts,amssymb,amsmath,amsthm}
\usepackage{bm}
\usepackage{physics}
\usepackage{xcolor}
\usepackage{hyperref}
\usepackage{orcidlink}
\usepackage[capitalize]{cleveref}
\usepackage{pifont}

\newtheorem{theorem}{Theorem}

\newcommand{\one}{\mathbb{I}}
\newcommand{\Hil}{\mathcal{H}}
\DeclareMathOperator{\supp}{supp}

\newcommand{\past}[1]{\overleftarrow{#1}}
\newcommand{\fut}[1]{\overrightarrow{#1}}
\newcommand{\both}[1]{\overleftrightarrow{#1}}
\newcommand{\fil}{_{\mathrm F}}
\newcommand{\rfil}{_{\mathrm R}}
\newcommand{\sm}{_{\mathrm S}}
\newcommand{\bx}{\mathbf{x}}
\newcommand{\by}{\mathbf{y}}

\newcommand{\xmark}{\ding{55}}

\definecolor{nblue}{rgb}{0.06,0.30,0.73}
\hypersetup{colorlinks=true,linkcolor=nblue,citecolor=nblue,urlcolor=nblue}

\begin{document}

\title{Unifying Quantum Smoothing Theories with Extended Retrodiction}

\author{Mingxuan Liu\orcidlink{0009-0009-4361-2641}}
\email{liu.mingxuan@u.nus.edu}
\affiliation{Centre for Quantum Technologies, National University of Singapore, Singapore 117543, Singapore}

\author{Ge Bai\orcidlink{0000-0002-6814-8840}}
\email{gebai@hkust-gz.edu.cn}
\affiliation{Thrust of Artificial Intelligence, Information Hub, The Hong Kong University of Science and Technology (Guangzhou), Guangzhou 511453, China}

\author{Valerio Scarani\orcidlink{0000-0001-5594-5616}}
\email{physv@nus.edu.sg}
\affiliation{Centre for Quantum Technologies, National University of Singapore, Singapore 117543, Singapore}
\affiliation{Department of Physics, National University of Singapore, Singapore 117542, Singapore}
\date{\today}

\begin{abstract}
Estimating the state of an open quantum system monitored over time requires incorporating information from past measurements (filtering) and, for improved accuracy, also from future measurements (smoothing). While classical smoothing is well understood within a Bayesian framework, its quantum generalization has been challenging, leading to distinct and seemingly incompatible approaches. In this work, we demonstrate that quantum state smoothing hinges on a uniquely quantum feature: the fundamental dependence of retrodiction on prior correlations. We introduce auxiliary systems into the prior belief to capture correlations formed during preparation and evolution and develop a comprehensive framework for quantum state smoothing based on extended Bayesian retrodiction. This framework identifies all previous approaches as different choices of the extended prior, and naturally extends it to other choices that have not been considered before. We also give an information-theoretic characterization of the choices of prior, in terms of the average entropy of the smoothed states. Our results establish quantum state smoothing as a fundamentally retrodictive process just like classical smoothing, with proper quantum features clearly identified.
 
\end{abstract}
\maketitle

\section{Introduction}
Classical estimation theory has three estimates for the state of dynamic open systems at time $t$ under continuous measurements: \emph{filtering}, or \emph{prediction}, takes into account only the information prior to $t$ \cite{Kalman60,Jaz07}; \emph{retrofiltering} takes into account only the information posterior to $t$; \emph{smoothing} takes into account both prior and posterior information \cite{Rauch63,Sarkka13}. Sometimes, the second and third techniques are not distinguished and are both recognized as \emph{retrodiction} because they can both be realized from Bayesian retrodiction, with prior chosen as the uniform distribution or the filtered state, respectively. In quantum information processing, the analogs of filtering and retrofiltering are straightforward, while quantum smoothing theory has been under debate for more than a decade~\cite{Watanabe55,Watanabe56,ABL64,Tsa-PRL09,Tsa-PRA09, CDJ13,  GJM13, GamMol14, Web14, GueWis15, CDJ15, Ohki15, CGLW21, LiuLaverick2025smoothing}.

The earliest attempts~\cite{Watanabe55,Watanabe56,ABL64}, although not formulated in the context of smoothing, are now known as the \emph{two-state vector formalism}, combining information from both pre- and post-selected states to describe the present state of a quantum system. Building upon this, Aharonov et al. introduced the concept of \emph{weak value} for intermediate measurements \cite{AAV88}. A subsequent and more general framework is the \emph{past quantum state}~\cite{GJM13}, which aims to incorporate future measurement information into estimating an unknown result of a quantum measurement, along the lines of smoothing, and has been successfully applied to several experimental settings \cite{TanMol15,BaoMol20b,BaoMol20a}. However, the ``smoothed'' objects here are not represented by standard density operators. 

The first method to define valid quantum states that are conditioned on both the prior and posterior measurement information was proposed by Guevara and Wiseman \cite{GueWis15}. In a nutshell, this method postulates that there is a ``complete'' measurement record, which defines the \emph{true} state that the system is left in. This record is distributed between Alice and Bob, and the quantum state smoothing problem becomes the question of how well, say, Alice can know the true state given the knowledge of her past and future measurement record. The problem has then been reduced to classical smoothing and can be computed accordingly. However, in many practical scenarios, ``Bob'' is just an anthropomorphism for the unmonitored part of the environment. In those cases, one has to assume a particular unraveling of the evolution, which raises concerns about the validity of the method~\cite{chantasri2025quantum}. 

Another approach to smoothing that always yields a valid state was recently obtained by adapting the procedure of how classical smoothing can be derived from retrodiction \cite{LiuLaverick2025smoothing}. The conditional Petz recovery map is used as the quantum Bayesian retrodiction, with the filtered state as prior. The resulting state, termed the Petz-Fuchs smoothed state due to its alignment with Fuchs' analysis \cite{Fuchs03}, is entirely independent of different unravelings of the evolution and thus requires no assumptions about a hypothetical observer Bob. Nevertheless, Bob may exist after all: the relation of this retrodictive approach with the Guevara-Wiseman smoothing was left open outside the case of pure initial states. 

In this work, we show that the quantum recipe for smoothing depends on the knowledge about correlated systems, which is a purely quantum feature absent in classical inference. Specifically, the recipe must include what is known about the ancilla involved in the preparation procedure (noticed for retrodiction in \cite{liu2026properimproper}), and about the environment that induces the open evolution. We develop a unified retrodictive framework that encompasses all forms of knowledge (e.g., agnosticism, classical correlation, quantum correlation), and identify those that underlie the Petz-Fuchs and the Guevara-Wiseman definitions as special cases. This perspective clarifies the role of priors in quantum retrodiction and places quantum smoothing in closer analogy to its classical counterpart. 

The remainder of this paper is organized as follows. In Section~\ref{sec:classical}, we review the elements of classical state inference, including filtering, retrofiltering, and smoothing. In particular, we show that the classical smoothed state can be obtained as a Bayesian retrodiction, derived by propagating the measurement record backward in time according to Bayes’ rule. Section~\ref{sec:quantum} recalls the corresponding quantum state filtering, retrofiltering, outlines the previously proposed desiderata for quantum smoothed state with minor modifications, and summarizes the two existing definitions of quantum smoothed state. In Section~\ref{sec:framework}, we develop our retrodictive framework from scratch and define the central result of this paper, the generalized smoothed state. We demonstrate how the earlier proposals arise as special cases. We then apply the formalism to more general scenarios. In Section~\ref{sec:entropy}, we establish lower and upper bounds on the average entropy of generalized smoothed states. Yet, we also prove that no future measurement-independent quantifier can universally order them. Section~\ref{sec:implication&connection} discusses the relation of our formalism with parameter estimation and counterfactual reasoning. Finally, Section~\ref{sec:conclusion} concludes with a summary of our results and open questions.

\section{Review: Classical state inference}\label{sec:classical}

\subsection{Set-up}
Consider an open classical system described by a collection of variables $\{x_i\}$, which we collect into a vector $\bx$. Because of interactions with an external environment, the dynamics of $\bx$ are stochastic. Thus, at any given time $t$, the system is described not by a single configuration but by a probability distribution. We denote the probability of the system being in state $\bx$ at time $t$ as $\wp(\bx;t)$. For generic probabilities other than the state of the system, we will keep using the usual notation $p$.  For clarity, we restrict our attention to discrete variables $\bx\in\mathbb{X}$, where $\mathbb{X}$ is finite or countably infinite.

Assuming Markovian dynamics, the evolution of $\wp(\bx;t)$ is governed by some forms of partial differential equation, e.g., the Fokker-Planck equation. In this work, we discretize the evolution from time $t_0$ to $T$ into an integer number of intervals, each of length $\dd t$. The dynamics at a time $t$ can be described by a stochastic matrix $D$, where $D(\bx;t+\dd t|\bx';t)$ describes the transition probability from $\bx'$ to $\bx$ and satisfies the completeness relation $\sum_{\bx\in{\mathbb X}}D(\bx;t+\dd t|\bx';t)=1$.

To estimate the state of the system without disturbing its dynamics, one can perform a sequence of non-disturbing measurements. At each time step $\dd t$, a measurement outcome $\by_t \in \mathbb{Y}$ is obtained. The entire measurement record between $t_0$ and $T$ is denoted by
\[
\both{{\bf O}}:=\{\by_\tau:\tau\in\{t_0,t_0+\dd t,\dots, T-\dd t\}\}.
\]
The inference task consists of estimating the state of the system at some intermediate time $t$, conditioned on different parts of this measurement record. The portion of the record prior to $t$ is denoted by
\[
\past{{\bf O}}_t:=\{\by_\tau:\tau\in\{t_0,t_0+\dd t,\dots, t-\dd t\}\},
\]
and the portion after $t$ by
\[
\fut{{\bf O}}_t:=\{\by_\tau:\tau\in\{t,t+\dd t,\dots, T-\dd t\}\}.
\]
This has naturally led to three distinct inference techniques: filtering (past only), retrofiltering (future only), and smoothing (both past and future).

\subsection{Filtering and Retrofiltering}
Suppose a measurement is performed at time $t$ and an outcome $\textbf{y}_t \in {\mathbb Y}$ is observed. The state of system at time $t$, $\wp(\bx;t)$, should be updated by conditioning on the information obtained from measurement. Bayes' theorem tells us
\begin{equation}\label{eq:bayesrule}
\wp(\bx;t|\textbf{y}_t)=\frac{p(\textbf{y}_t|\bx;t)\wp(\bx;t)}{p(\textbf{y}_t)}.
\end{equation}

Combining this together with the transition probability $D(\bx;t+\dd t|\bx';t)$, we should describe the post-measurement state as
\begin{align}
     \wp(\bx;t+\dd t|\textbf{y}_t)&=\sum_{\bx'\in\mathbb{X}}D(\bx;t+\dd t|\bx';t)\wp(\bx';t|\by_t) \nonumber\\
     &=\frac{\sum_{\bx'\in{\mathbb X}}\varphi_{\textbf{y}_t}(\bx;t+\dd t|\bx';t)\wp(\bx';t)}{p(\textbf{y}_t)},\label{eq:update}
\end{align}
where we denote the overall conditional dynamical map by $\varphi_{\textbf{y}_t}(\bx;t+\dd t|\bx';t):=D(\bx;t+\dd t|\bx';t)p(\textbf{y}_t|\bx';t)$. The family $\{\varphi_{\by_t}\}_{\by_t\in\mathbb{Y}}$ is the classical analog of a quantum instrument: summing over all $\by_t$ recovers the unconditional dynamics $D$.  

Filtering is an estimation technique that estimates the state of interest based on past information, i.e., $\wp\fil(\bx;t):=\wp(\bx;t|\past{{\bf O}}_t)$. If taking the filtered state at time $t$ as the prior state of \cref{eq:update}, we thus obtain the filtered state at the next time step:
\begin{equation}\label{eq:recurfil}
\wp\fil(\bx;t+\dd t)=\frac{\sum_{\bx'\in{\mathbb X}}\varphi_{\textbf{y}_t}(\bx;t+\dd t|\bx';t)\wp\fil(\bx';t)}{p(\textbf{y}_t|\past{{\bf O}}_t)}.
\end{equation}
Consequently, with the initial state $\wp_0(\bx):=\wp(\bx;t_0)$, one can derive the general formula for classical filtered state:
\begin{equation}
\wp\fil(\bx;t)=\frac{\sum_{\bx'\in{\mathbb X}}\varphi_{\past{\bf O}_t}(\bx|\bx')\wp_0(\bx')}{p(\past{{\bf O}}_t)},
\end{equation}
where $\varphi_{\past{\bf O}_t}$ is the composition of all conditional dynamical maps from time $t_0$ to $t-\dd t$: $\varphi_{\past{\bf O}_t}=\varphi_{\textbf{y}_{t-\dd t}}\varphi_{\textbf{y}_{t-2\dd t}}\cdots\varphi_{\textbf{y}_{t_0}}$.

In contrast, retrofiltering is an estimation technique that estimates the state of interest based on future information only. Retrofiltered effect, not necessarily to be a state, is defined as $E\rfil(\bx;t):=p(\fut{\bf O}_t|\bx;t)$. Noticing the consistency relation $\sum_{x\in{\mathbb X}}{\wp\fil(\bx;t) E\rfil(\bx;t)}\cdot p(\past{\bf O}_t)=p(\both{\bf O})$ and \cref{eq:recurfil}, retrofiltered effect evolves backwards according to:
\begin{equation}\label{eq:recurrfil}
E\rfil(\bx;t)=\sum_{\bx'\in\mathbb{X}}\varphi_{\textbf{y}_t}(\bx';t+\dd t|\bx;t)E\rfil(\bx';t+\dd t).
\end{equation}
We can likewise write down the general formula for classical retrofiltered effect:
\begin{equation}
E\rfil(\bx;t)=\sum_{\bx'\in{\mathbb X}}\varphi_{\fut{\bf O}_t}(\bx'|\bx)
\end{equation}
where an uninformative final effect $E\rfil(\bx';T)=1$ for all $\bx'\in\mathbb{X}$ is included implicitly and $\varphi_{\fut{\bf O}_t}$ is the composition of all conditional dynamical maps from time $t$ to $T-\dd t$: $\varphi_{\fut{\bf O}_t}=\varphi_{\textbf{y}_{T-\dd t}}\varphi_{\textbf{y}_{T-2\dd t}}\cdots\varphi_{\textbf{y}_{t}}$.

\subsection{Smoothing} 
As a combination of above two estimation techniques, smoothing takes information from both the past and future into account to estimate the state at a specific moment. Smoothed state is analogously defined as $\wp\sm(\bx;t):=\wp(\bx;t|\both{{\bf O}})$. Adapting above result, it is not difficult to show that
\begin{equation}
\wp\sm(\bx;t) \propto \wp\fil(\bx;t)E\rfil(\bx;t),
\end{equation}
which is because
\begin{align*}
    \wp(\bx;t|\both{{\bf O}})=&\frac{p(\both{\bf O}|\bx;t)\wp(\bx;t)}{p(\both{\bf O})}\\=&\frac{p(\fut{\bf O}_t|\past{\bf O}_t,\bx;t)p(\past{\bf O}_t|\bx;t)\wp(\bx;t)}{p(\both{\bf O})}\\=&\frac{p(\fut{\bf O}_t|\bx;t)\wp(\bx;t|\past{\bf O}_t)p(\past{\bf O}_t)}{p(\both{\bf O})}\\=&\frac{\wp\fil(\bx;t)E\rfil(\bx;t)}{p(\fut{\bf O}_t|\past{\bf O}_t)}.
\end{align*}
Bayes' rule and the assumption of Markov process are used in this derivation. Equivalently, we may express this in a retrodictive form
\begin{align}\label{eq:c_retrodiction}
\wp\sm(\bx;t) &= \sum_{\fut{\bf O'}_t}\frac{p(\fut{\bf O'}_t|\bx;t)\wp(\bx;t|\past{\bf O}_t)}{p(\fut{\bf O'}_t|\past{\bf O}_t)} \delta_{\fut{\bf O'}_t,\fut{\bf O}_t} \nonumber\\
&= \sum_{\fut{\bf O'}_t} \mathcal{R}(\bx;t|\fut{\bf O'}_t)\delta_{\fut{\bf O'}_t,\fut{\bf O}_t},
\end{align}
where $\delta_{\fut{\bf O'}_t,\fut{\bf O}_t}$ is the Kronecker delta. The object $\mathcal{R}(\bx;t|\fut{\bf O'}_t):={p(\fut{\bf O'}_t|\bx;t)\wp(\bx;t|\past{\bf O}_t)}/{p(\fut{\bf O'}_t|\past{\bf O}_t)}$ can be seen as a ``reverse process'' and used to retrodict on observed evidence probability distributions, according to Jeffrey’s probability kinematics \cite{jeffrey} and Pearl’s virtual evidence method \cite{pearl}.

\section{Review: Quantum state inference}\label{sec:quantum}
\subsection{Set-up}
We now switch focus to estimating the dynamics of open quantum systems. Due to the interactions with multiple external baths, the state of system, described by a density matrix $\rho$, evolves nonunitarily. Markovianity is also assumed here, and thus the time evolution of the state is governed by the master equation of Lindblad form~\cite{Lin76}. To monitor the system without involving much additional dynamics, one can extract information about the system by performing continuous measurement directly onto the baths, namely that after each infinitesimal time interval $\dd t$, a generalized measurement is performed and a record is obtained. With access to information about the system at time $t_0$, e.g., initial state $\rho_0$, and its evolution process, e.g., measurement record $\bf O$, the agent (Alice) can estimate the state of the system throughout the entire process conditionally.

\subsection{Filtering and Retrofiltering}

Given measurement outcome prior to the estimation time, conventional quantum theory prescribes how to predict the quantum state at that time. It is also known as quantum filtering \cite{WisMil10}. We consider the case of inefficient measurement, where the observation is made with less-than-unit readout efficiency. Such a process can equivalently be viewed as a partially efficient measurement accompanied by dissipation, namely that the observer, Alice, monitors only part of the environment while the unmonitored baths induce additional decoherence. The corresponding conditional quantum operation can be written as $\Phi_{{\bf y}_t}(\bullet)=\sum_{k=1}M_{k|{\bf y}_t}\bullet M_{k|{\bf y}_t}^\dagger$, and the quantum {\em filtered state} evolves according to
\begin{equation}
\rho\fil(t+\dd t)=\frac{\Phi_{{\bf y}_t}\left(\rho\fil(t)\right)}{\Tr\left[\Phi_{{\bf y}_t}\left(\rho\fil(t)\right)\right]}.
\end{equation}
The collection $\{\Phi_{\by_t}\}$ constitutes a {\em quantum instrument} with $\sum_{\by_t} \Phi_{\by_t}^\dag (\one)=\one$; where the adjoint of any linear map $\Phi$ is defined by $\Tr[\Phi(X)^\dag ~ Y] = \Tr[X^\dag ~\Phi^\dag(Y)]$, for all operators $X$ and $Y$. This convention applies to all the other conditional quantum operations considered in this paper. We define the composite quantum operation for the interval $[0,t)$ conditioned on the record $\past{\bf O}_t$ of evenly spaced measurements as
$\Phi_{\past{\bf O}_t}:=\Phi_{{\bf y}_{t-\dd t}}\circ\Phi_{{\bf y}_{t-2\dd t}}\circ\cdots\circ\Phi_{{\bf y}_{t_0}}$. With this, the filtered state can be expressed as
\begin{equation}\label{eq:filtering}
\rho\fil(t)=\frac{\Phi_{\past{\bf O}_t}(\rho_0)}{\Tr[\Phi_{\past{\bf O}_t}(\rho_0)]}.
\end{equation}

The \emph{retrofiltered effect} has been introduced in \cite{GJM13,GueWis15}. It can be thought of as evolving backwards from an uninformative identity conditioned on the measurement record $\fut{\bf O}_t$:
\begin{equation}\label{eq:retrofiltering}
\hat{E}\rfil(t)=\Phi_{\fut{\bf O}_t}^\dagger \left(\one\right),
\end{equation}
where $\Phi_{\fut{\bf O}_t}^\dag$ is the adjoint of the conditional composite quantum operation $\Phi_{\fut{\bf O}_t}:=\Phi_{{\bf y}_{T-\dd t}}\circ\Phi_{{\bf y}_{T-2\dd t}}\circ\cdots\circ\Phi_{{\bf y}_{t}}$. After normalization, one can identify it as the output of the retrodiction map defined in \cref{sec:framework} when the prior state is chosen to be an uninformative maximally mixed state.

\subsection{Smoothing}
\label{sec:smoothing}

In classical smoothing theory, the smoothed state at time $t$ is a single, well-defined probability distribution $\wp\sm$ with strictly valid statistical properties conditioned on both past and future measurement records. For a quantum analog, we seek an estimation of quantum state, $\rho\sm$, that is conditioned on measurement record of the entire process as well. The following criteria are required (adjusted from Ref.~\cite{LCW-QS21}):

\underline{\bf Criterion 1:} {\em The theory should yield a physical quantum state $\rho\sm$ (positive semidefinite and unit trace), directly analogous to the classical smoothed state $\wp\sm$.}\\

\underline{\bf Criterion 2:} {\em The smoothed state should reduce to the filtered state when averaged over all possible future measurement records, given a fixed past information.}\\

\underline{\bf Criterion 3:} {\em In the limit where the agent's belief about the initial conditions, evolution and measurement processes can all be described probabilistically in a fixed basis, the smoothed quantum state should reduce to the classical smoothed distribution.}

Several proposals for quantum state smoothing have been introduced in the literature. We now recall two of them that satisfy all these desiderata.

\subsubsection{Guevara-Wiseman Smoothing}
\label{sec:GW-smoothing}

Guevara and Wiseman~\cite{GueWis15} framed quantum smoothing in terms of \emph{true quantum trajectories}. Besides Alice performing partially efficient measurements and recording the observed outcomes $\bf O$, assume the remainder of the environment is measured by a hypothetical second observer Bob, who obtains outcomes $\bf U$. The \emph{true state} of the system at time $t$ is defined as the filtered quantum state conditioned on both Alice’s and Bob’s past records, i.e., $\rho(t|\past{\bf O}_t, \past{\bf U}_t)=\Phi_{\past{\bf O}_t, \past{\bf U}_t}(\rho_0)/\Tr[\Phi_{\past{\bf O}_t, \past{\bf U}_t}(\rho_0)]$. 

From Alice’s perspective, the usual filtered state $\rho\fil(t)$ is then a classical mixture over possible true states given only her past record $\past{\bf O}_t$. Guevara and Wiseman defined the smoothed state $\rho^{\rm GW}\sm$ as a mixture over true states conditioned on both Alice’s past and future records $\both{\bf O}$:
\begin{equation}\label{eq:GW-smoothing}
    \rho^{\rm GW}\sm(t)=\sum_{\past{\bf U}_t}p(\past{\bf U}_t|\both{\bf O})\rho(t|\past{\bf O}_t, \past{\bf U}_t).
\end{equation}

However, the approach relies on the assumption about Bob’s measurement. In physical settings where Bob does not exist, or where no information about his measurement is available, one must posit a particular unraveling of the evolution, which introduces ambiguity. Moreover, computing $\rho^{\rm GW}\sm$ generally requires sampling over many possible trajectories, making the approach computationally demanding.

\subsubsection{Petz-Fuchs Smoothing}
\label{sec:PF-smoothing}

A conceptually different route is based on quantum Bayesian retrodiction. Using Petz recovery map~\cite{Petz1986} conditionally, a so-called Petz-Fuchs smoothed state is defined analogously to classical smoothing \cite{LiuLaverick2025smoothing}. Propagating the filtered state at final time $T$ backward through conditional Petz recovery map yields
\begin{equation}\label{eq:PF-smoothing}
    \rho^{\rm PF}\sm(t) = \frac{\sqrt{\rho\fil(t)}\, \hat{E}\rfil(t)\, \sqrt{\rho\fil(t)}}{\Tr\!\left[\rho\fil(t) \hat{E}\rfil(t)\right]},
\end{equation}
where $\rho\fil(t)$ is the filtered state and $\hat{E}\rfil(t)$ is the retrofiltered effect. Both are under the definitions in the previous subsection. Apparently, the Petz-Fuchs smoothed state incorporates both past and future information in the same way as classical smoothed state---product of filtered state and retrofiltered effect. This turns out also to be a special application of the quantum Bayes' rule proposed by Fuchs~\cite{Fuchs03}.

Unlike the Guevara-Wiseman construction, the Petz-Fuchs state does not require an assumption on the hypothetical observer Bob or an unraveling of the evolution. In addition, it admits efficient computation through closed expressions. However, its interpretation as conditioning on future records is less direct than in the trajectory picture, making its applicability obscure. 

\bigskip
Taken together, these two approaches illustrate the central tension in quantum smoothing: two distinct definitions each have advantages over the other, but there is no consistent way to understand both. In the next section, we present an extended retrodictive framework that reconciles these perspectives.

\section{Retrodictive Framework for Quantum Smoothing}\label{sec:framework}
\subsection{General Formalism}
From \cref{eq:c_retrodiction}, we see that classical smoothing can be interpreted as propagating the evidence through a reverse process derived from classical Bayesian retrodiction. Petz recovery map~\cite{Petz1986} is a canonical choice for a quantum version of the reverse process, with its uniqueness established in various works \cite{clive24tabletop,SS23,PB23,PF23,Scandi25QFIPetz}. In the case of measurement processes (the focus of this paper), both the posterior state and the new evidence are classical, and therefore commute. As a consequence, the resulting update rule satisfies the minimum change principle \cite{Ge2025MCP,kuang2025quantummeasurementretrodictionentropic}, in direct analogy with classical Bayes’ rule.
For a forward channel $\mathcal{E}$ from system $Q$ to system $T$ and a prior belief $\gamma \in S(\Hil_Q)$ about the state of system $Q$, we denote by $\mathcal{R}^{\mathcal{E},\gamma}$ the corresponding Petz recovery map.
If one obtains an evidence $\sigma \in S(\Hil_T)$, they can apply the Petz recovery map to it and update their belief about $Q$ as
\begin{align} \label{eq:Petz}
    \mathcal{R}^{\mathcal{E},\gamma}(\sigma) := \sqrt{\gamma}\mathcal{E}^\dagger\left(\mathcal{E}(\gamma)^{-\frac{1}{2}}\sigma\mathcal{E}(\gamma)^{-\frac{1}{2}}\right) \sqrt{\gamma}.
\end{align}
A mixed quantum state can be interpreted in two distinct ways. It may describe classical ignorance about which definite quantum state the system is in within a specified ensemble---a \emph{proper mixture}; or it may arise from entanglement with another system that has been traced out---an \emph{improper mixture} \cite{d1976conceptual}. It has been shown that they correspond to different priors for quantum state retrodiction \cite{liu2026properimproper}, which gives rise to the need of extending retrodiction to a larger Hilbert space that incorporates an auxiliary system $A$, $\Hil_Q\otimes\Hil_A$, where all known correlations with the system $Q$ are included. Accordingly, the prior belief should be a density operator $\Gamma$ lying in $S(\Hil_Q\otimes \Hil_A)$. The prior-extended retrodiction map reads
\begin{align}\label{eq:prior-ext_Petz}
    &\mathcal{R}_{\rm ext}^{\mathcal{E},\Gamma}(\sigma) := (\Tr_A \circ \mathcal{R}^{\mathcal{E} \otimes \Tr, \Gamma})(\sigma)\\=&\Tr_A\left[\sqrt{\Gamma}\left(\mathcal{E}^\dag(\mathcal{E}(\gamma)^{-\frac{1}{2}}\sigma\mathcal{E}(\gamma)^{-\frac{1}{2}}) \otimes\one_A\right)\sqrt{\Gamma}\right] \nonumber
\end{align}
where $\gamma := \Tr_A[\Gamma]$.

\begin{figure}
    \centering
    \includegraphics[width=1\linewidth]{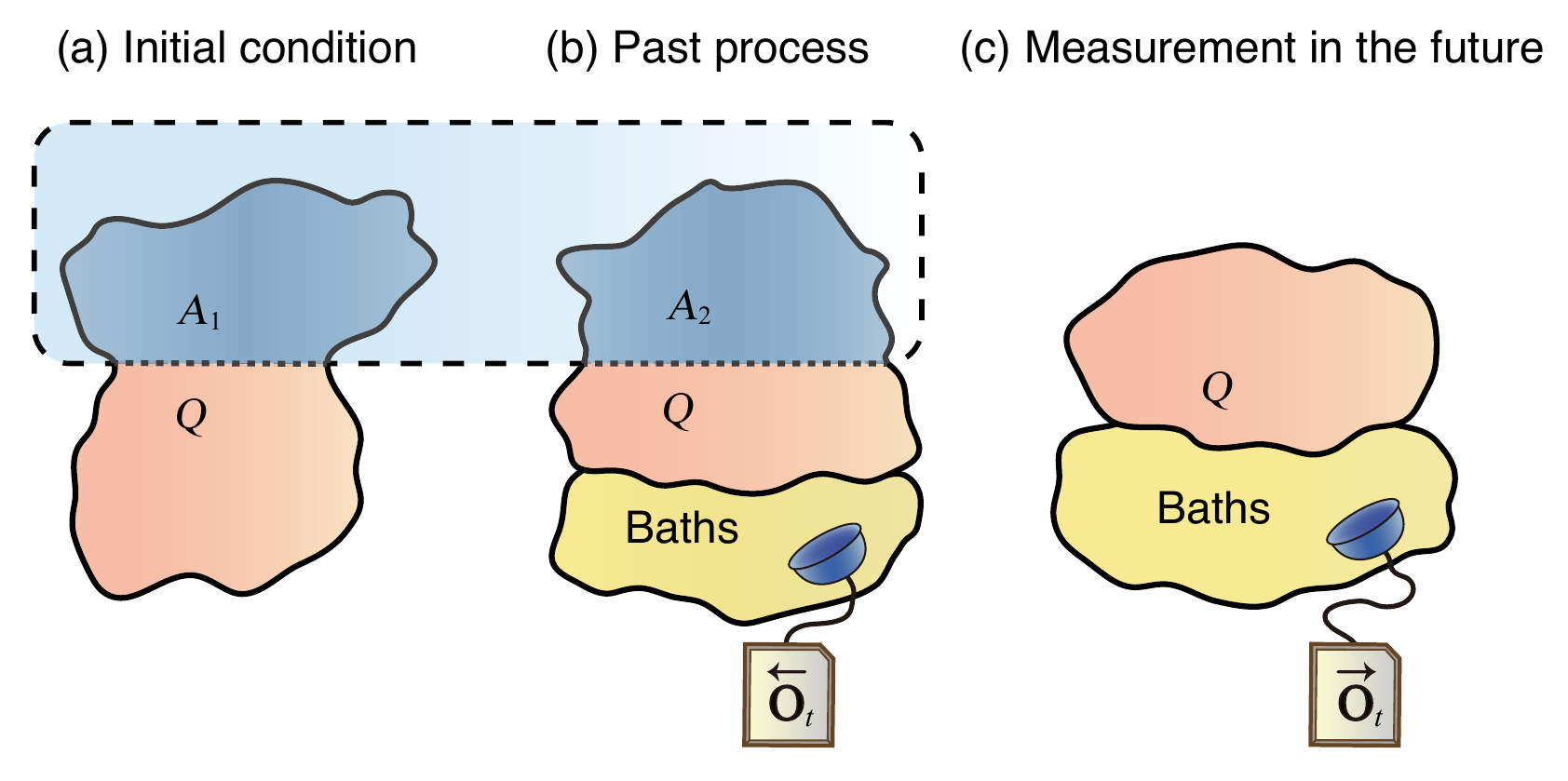}
    \caption{Schematic illustration of the information structure underlying the generalized quantum smoothing formalism. (a) The initial condition may include correlations between the system $Q$ and an auxiliary subsystem $A_1$, representing knowledge about the system’s preparation. (b) The past process involves the interaction of $Q$ with several baths, part of which may be monitored by Alice, producing the observed record $\past{\bf O}_t$. The auxiliary subsystem $A_2$ captures possible correlations with unmonitored baths. Together, $A_1$ and $A_2$ constitute the total auxiliary system $A$, forming the extended prior described by the filtered global state $\varrho\fil(t)$. (c) The future measurement on the monitored bath yields the record $\fut{\bf O}_t$, which provides the evidence that Alice retrodicts on to construct the smoothed state $\rho\sm(t)$.}
    \label{fig:q_smoothing}
\end{figure}

We now apply this idea to smoothing. To this end, we introduce a new concept, the {\em filtered global state}, denoted by $\varrho\fil(t)$ (recall that the filtered state is $\rho\fil(t)$). It is the state of the system $Q$ and of an auxiliary system $A$ that encodes information about the correlations with system $Q$ based on Alice's knowledge prior to time $t$. Note that it has to be consistent with the previous definition; basically, if we care about the state of system $Q$ locally, the filtered global state should give the same answer as the filtered state does, i.e. $\rho\fil(t) = \Tr_A[\varrho\fil(t)]$. We can describe the continuous measurement process from time $t$ to $T$, for any possible observed measurement record $\fut{\bf O}_t$, by the quantum--classical channel 
\begin{equation}
    {\cal E}_{t\to T}[\bullet]=\sum_{\fut{\bf O}_t}\Tr\left[\Phi_{\fut{\bf O}_t}[\bullet]
    \right]\ketbra{\fut{\bf O}_t}{\fut{\bf O}_t},
\end{equation}
where $\ket{\fut{\bf O}_t}$ represents a classical state and thus forms an orthonormal basis, namely that $\braket{\fut{\bf O}_t}{\fut{\bf O}'_t}=\delta_{\fut{\bf O}_t,\fut{\bf O}'_t}$. Applying filtered global state $\varrho\fil(t)$ as the extended prior to construct the retrodiction map of this measurement process, if the evidence observed by Alice in period $[t,T)$ is $\ketbra{\fut{\bf O}_t}{\fut{\bf O}_t}$, we have the {\em generalized quantum smoothed state}
\begin{equation}\label{eq:smoothig}
\begin{split}
    \rho\sm(t)&=\mathcal{R}_{\rm ext}^{{\cal E}_{t\to T},\varrho\fil(t)}(\ketbra{\fut{\bf O}_t}{\fut{\bf O}_t})\\&=\frac{\Tr_A\left[\sqrt{\varrho\fil(t)}\left(\hat{E}\rfil(t)\otimes\one_A\right)\sqrt{\varrho\fil(t)}\right]}{\Tr\left[\rho\fil(t)\hat{E}\rfil(t)\right]}.
\end{split}
\end{equation}
This definition satisfies all the desired properties and thus yields a valid quantum smoothed state (see Appendix~\ref{app:proof_smooth} for the full proof). This formula depends not only on the entire measurement record $\both{\bf O}$, but also on Alice's knowledge about the existing correlations with system $Q$ prior to time $t$. We will demonstrate that when the filtered global state is appropriately constructed from prior information, we can unify the aforementioned Guevara-Wiseman smoothing and Petz-Fuchs smoothing with \cref{eq:smoothig}. Of course, our formalism does not encompass just those two cases, but describes smoothing under any other scenario of information (see \cref{fig:q_smoothing}).

\subsection{Recovering Petz-Fuchs Smoothing}
When nothing but the density operator of the system $Q$ is known to Alice or, say, all the other information that Alice has access to is uncorrelated with the system $Q$, Alice is agnostic to the unknown environment. The auxiliary system $A$ becomes a trivial 1-dimensional space, namely that $\varrho^{\rm PF}\fil(t)=\rho\fil(t)$; and \cref{eq:smoothig} reduces to \cref{eq:PF-smoothing}.

\subsection{Recovering Guevara-Wiseman smoothing}
The attempt of expressing Guevara-Wiseman smoothed state in terms of Petz recovery map was not successful in \cite{LiuLaverick2025smoothing}. Here, we provide a new perspective provided by prior extension. Suppose Alice knows the initial state $\rho_0$ with certainty and knows what measurement Bob performs. Then, considering smoothed state as a mixture of all possible {\em true} states, $\rho(t|\past{\bf O}_t, \past{\bf U}_t)$, is equivalent to setting the filtered global state as 
\begin{equation}
    \varrho^{\rm GW}\fil(t)=\sum_{\past{\bf U}_t}\frac{(\Phi_{\past{\bf O}_t,\past{\bf U}_t} \otimes \textrm{id})(\ketbra{\Psi}{\Psi}_{QA_1})}{\Tr[\Phi_{\past{\bf O}_t}(\rho_0)]}\otimes \ketbra{\past{\bf U}_t}{\past{\bf U}_t}_{A_2}
\end{equation}
where $\ketbra{\Psi}{\Psi}_{QA_1}$ is any purification of $\rho_0$. Using the purification (i.e., improper mixture) reflects a certain belief in system $Q$ being in state $\rho_0$ and thus a refusal to update \cite{liu2026properimproper}. $A_2$ is a hypothetical classical register with $\braket{\past{\bf U}_t}{\past{\bf U}'_t}=\delta_{\past{\bf U}_t,\past{\bf U}'_t}$, which is classically correlated with system $Q$, reflecting Alice's knowledge about Bob's measurement. Note, again, consistency is satisfied, $\Tr_A\left[\varrho^{\rm GW}\fil(t)\right]=\rho\fil(t)$. Plugging it into \cref{eq:smoothig} gives back the original expression \cref{eq:GW-smoothing}:
\begin{widetext}
\begin{equation}
\begin{split}
    &\frac{1}{\Tr\left[\rho\fil(t)\hat{E}\rfil(t)\right]}{\Tr_A\left[\sqrt{\varrho^{\rm GW}\fil(t)}\left(\hat{E}\rfil(t)\otimes\one_A\right)\sqrt{\varrho^{\rm GW}\fil(t)}\right]}\\
    =&\frac{1}{\Tr[\Phi_{\past{\bf O}_t}(\rho_0)]\Tr\left[\rho\fil(t)\hat{E}\rfil(t)\right]}\sum_{\past{\bf U}_t}\Tr_{A_1}\left[\sqrt{(\Phi_{\past{\bf O}_t,\past{\bf U}_t} \otimes \textrm{id})(\ketbra{\Psi}{\Psi})}\left(\hat{E}\rfil(t)\otimes\one_{A_1}\right)\sqrt{(\Phi_{\past{\bf O}_t,\past{\bf U}_t} \otimes \textrm{id})(\ketbra{\Psi}{\Psi})}\right]\\
    =&\frac{1}{\Tr[\Phi_{\both{\bf O}}(\rho_0)]}\sum_{\past{\bf U}_t} \frac{\Tr\left[\left(\Phi_{\fut{\bf O}_t}^\dag(\one_Q)\otimes\one_{A_1}\right)(\Phi_{\past{\bf O}_t,\past{\bf U}_t} \otimes \textrm{id})(\ketbra{\Psi}{\Psi})\right]}{\Tr\left[(\Phi_{\past{\bf O}_t,\past{\bf U}_t} \otimes \textrm{id})(\ketbra{\Psi}{\Psi})\right]} \Tr_{A_1}\left[(\Phi_{\past{\bf O}_t,\past{\bf U}_t} \otimes \textrm{id})(\ketbra{\Psi}{\Psi})\right]\\
    =&\frac{1}{\Tr[\Phi_{\both{\bf O}}(\rho_0)]}\sum_{\past{\bf U}_t} {\Tr\left[\Phi_{\fut{\bf O}_t}^\dag(\one_Q) \Phi_{\past{\bf O}_t,\past{\bf U}_t} (\rho_0)\right]}\frac{\Phi_{\past{\bf O}_t,\past{\bf U}_t}(\rho_0)}{\Tr\left[\Phi_{\past{\bf O}_t,\past{\bf U}_t}(\rho_0)\right]}=\sum_{\past{\bf U}_t} \frac{p({\both{\bf O},\past{\bf U}_t})}{p(\both{\bf O})}\rho(t|{\past{\bf O}_t,\past{\bf U}_t})=\rho^{\rm GW}\sm(t).
\end{split}
\end{equation}
\end{widetext}
In the second line, we trace out the classical register $A_2$. For the subsequent equality, we used the definitions of the filtered state and retrofiltered effect, along with the fact that each $\Phi_{\past{\bf O}_t,\past{\bf U}_t}$ has Kraus rank one. In the last line, we use that the trace of a state after a conditional operation equals the probability associated with that condition.

\subsection{Beyond Existing Definitions}
Before proceeding, we need to emphasize that subjectivity plays a role in Bayesian reasoning. The extension of prior introduces an extra degree of freedom to make it happen as compared to the classical case, reflecting the agent’s beliefs about unobserved correlations. In what follows, we discuss several representative scenarios and provide the corresponding reasonable choices of extended priors from our perspective as the filtered global state. The use of auxiliary systems in each case is summarized in \cref{tab:smoothed_states}.

\subsubsection{Guevara-Wiseman variant}
This was vaguely mentioned in \cite{LiuLaverick2025smoothing} but not classified as a valid smoothed state. Similar to the Guevara-Wiseman scenario, here we also assume the remainder of baths are monitored by Bob and Alice knows what measurement Bob performs. However, Alice is agnostic to the initial state $\rho_0$. As a consequence, from Alice's perspective, after performing measurements continuously until time $t$, her filtered global state is
\begin{equation}
\varrho\fil(t)=\sum_{\past{\bf U}_t} \frac{\Phi_{\past{\bf O}_t,\past{\bf U}_t}(\rho_0)}{\Tr[\Phi_{\past{\bf O}_t}(\rho_0)]}\otimes\ketbra{\past{\bf U}_t}{\past{\bf U}_t}_A,
\end{equation}
which yields the smoothed state
\begin{equation}\label{eq:GW-smoothing-variant}
    \rho\sm(t)
    =\sum_{\past{\bf U}_t} \frac{\sqrt{\Phi_{\past{\bf O}_t,\past{\bf U}_t}(\rho_0)}\hat{E}\rfil(t)\sqrt{\Phi_{\past{\bf O}_t,\past{\bf U}_t}(\rho_0)}}{\Tr\left[\Phi_{\past{\bf O}_t}(\rho_0)\hat{E}\rfil(t)\right]}.
\end{equation}

Similar to Guevara-Wiseman smoothing, we observe that \cref{eq:GW-smoothing-variant} is also the average over the probability distribution $p(\past{\bf U}_t|\both{\bf O})$. However, the quantum states being averaged differ due to agents' distinct beliefs about the initial state. They coincide if $\rho_0$ is already a pure state.

\subsubsection{Petz-Fuchs variant}
It is usually the case that an experimentalist has control of the initial state while knowing little about the environment. In this case, Alice may pick a certain initial belief and choose not to impose any hypothetical constraints on the evolution process. Naturally, her filtered global state turns out to be
\begin{equation}
    \varrho\fil(t)=\frac{(\Phi_{\past{\bf O}_t} \otimes \textrm{id})(\ketbra{\Psi}{\Psi}_{QA})}{\Tr[\Phi_{\past{\bf O}_t}(\rho_0)]},
\end{equation}
where $\ketbra{\Psi}{\Psi}_{QA}$ is a purification of the initial state $\rho_0$. Plugging into \cref{eq:smoothig} yields the corresponding smoothed state. Though the expression cannot be further simplified, we want to mention that the smoothed state stays invariant under any local isometry to the auxiliary system $A$, since any purification extension represents equivalent certain belief \cite{liu2026properimproper}.

\subsubsection{Agent in the Church of the Larger Hilbert Space}
The most extreme case for prior extension is when the agent stays in the \emph{Church of the Larger Hilbert Space} (a term coined by John Smolin and explicated in Ref.~\cite{Gottesman2000churchoflargerhilbertspace}). It is to say that Alice believes the quantum system $Q$ is always a marginal of a pure entangled state. Or, she is certain that there is no adversary, say Bob, stealing information about system $Q$. Therefore, her filtered global state should be a purification of the filtered state, i.e., $\varrho\fil^{\rm CLHS}(t)=\ketbra{\Psi}{\Psi}_{QA}$. Plugging into \cref{eq:smoothig} gives Alice's smoothed state,
\begin{equation}
    \rho\sm^{\rm CLHS}(t)=\Tr_A(\ketbra{\Psi}{\Psi}_{QA})=\rho\fil(t).
\end{equation}
As one would expect, since Alice assigns certainty to her prior, her belief about the state of system $Q$ is unaffected by the arrival of any further measurement record.

\begin{table}[h]
\centering
\caption{
Summary of the five types of quantum smoothed states discussed in this work. Each scenario is characterized by its filtered global state $\varrho\fil(t)$, distinguished by what correlation information in the initial condition and through the past process is encoded via the use of an auxiliary system. 
``\xmark'' indicates that no auxiliary system is employed, leaving the agent’s ignorance completely unspecified. 
``P'' denotes that the auxiliary system is used to represent a proper mixture, while ``I'' denotes that it represents an improper mixture.
}
\label{tab:smoothed_states}
\vspace{5pt}
\begin{tabular}{lcc}
\hline\hline
Scenario & Initial condition & Past process \\
\hline
Petz-Fuchs & \xmark & \xmark \\
Guevara-Wiseman & I & P \\
Guevara-Wiseman Variant & \xmark & P \\
Petz-Fuchs Variant & I & \xmark \\
CLHS & I & I \\
\hline\hline
\end{tabular}
\end{table}

\section{Entropic relation for generalized smoothed state}\label{sec:entropy}
\subsection{Bounds on average entropy of smoothed state}
Uncertainty is expected to reduce when conditioning on more information, regardless of whether the estimation is predictive or retrodictive. Average von Neumann entropy is a natural figure of merit for quantifying uncertainty. Because Criterion 2 is always satisfied by any valid definition of smoothing, the filtered state can be regarded as a mixture of smoothed states corresponding to all possible future measurement records. For Alice with observed past record $\past{\bf O}_t$ and filtered state $\rho\fil(t)$, once she decides her future measurement choice, the average entropy of smoothed state, $\bar{S}(\rho\sm(t)):=\mathbb{E}_{\fut{\bf O}_t}[S(\rho_{\text S|\fut{\bf O}_t}(t))]$, is bounded by 
\begin{equation}
    S(\rho\fil(t))-H(\fut{\bf O}_t)\le \bar{S}(\rho\sm(t))\le S(\rho\fil(t)),
\end{equation}
where $S$ and $H$ denote von Neumann entropy and Shannon entropy functions, respectively. Note that the conditioning on $\past{\bf O}_t$ is implicit for all terms. It can be easily obtained from standard quantum information theory \cite{nielsen2010qcqi}. Furthermore, a more specific lower bound for average entropy of Guvuara-Wiseman smoothed state was discussed in Ref.~\cite{Budini18b}. 

In this section, we investigate the average von Neumann entropy of our generalized smoothed states to understand how different choices of prior extension affect the behavior of quantum smoothing. Our result, \cref{thm:bounds}, shows that the Petz–Fuchs smoothed state achieves the minimal average von Neumann entropy, whereas the smoothed state for agents in the Church of the Larger Hilbert Space attains the maximal average, namely for any smoothed state $\rho_S(t)$, one has
\begin{equation}
    \bar{S}(\rho^{\rm PF}\sm(t))\le \bar{S}(\rho\sm(t))\le \bar{S}(\rho^{\rm CLHS}\sm(t)).
\end{equation}

\begin{theorem}[Lower and upper bounds on average entropy]\label{thm:bounds}
Let $\Hil_Q$ be a finite-dimensional system Hilbert space, $\gamma\in\mathcal S(\Hil_Q)$ a density operator. Consider a POVM $\{E_i\}_i$ on $\Hil_Q$. For any extension $\Gamma\in\mathcal S(\Hil_Q\otimes \Hil_A)$ satisfying $\Tr_A\Gamma=\gamma$, for each outcome $i$, the probability of obtaining this outcome by measuring $\gamma$ is
\begin{align}\label{eq:p-i}
    p_{E_i}:=\Tr\left[E_i\gamma\right].
\end{align}
The smoothed state corresponding to this outcome, by \cref{eq:smoothig}, is
\begin{align}\label{eq:rho-i-gamma}
    \rho_{E_i}^{\Gamma}:=\frac{\Tr_A\left[\sqrt{\Gamma}\,(E_i\otimes\one_A)\,\sqrt{\Gamma}\right]}{p_i}.
\end{align}
Consider the following extensions of the $\gamma$:
\begin{enumerate}
    \item Trivial extension, namely $\gamma$ itself. This, for outcome $i$, leads to the smoothed state
    \[    \rho_{E_i}^{\gamma}:=\frac{\sqrt{\gamma}\,E_i\,\sqrt{\gamma}}{p_i}\]
    \item A purification, namely a pure state $\ket{\Phi}\in\Hil_Q\otimes\Hil_A$ satisfying $\Tr_A[\ketbra{\Phi}{\Phi}] = \gamma$. This leads to the smoothed state equal to $\gamma$ independent of the outcome $i$.
\end{enumerate}

Then the $p$-weighted average von Neumann entropy satisfies
\begin{align}
\sum_i p_{E_i}\,S(\rho_{E_i}^\gamma)\ \le\ \sum_i p_{E_i}\,S(\rho_{E_i}^\Gamma) \ \le \ S(\gamma),
\end{align}
i.e. the choice $\Gamma=\gamma$ (trivial extension) minimizes the average entropy among all extensions with the same system marginal $\gamma$, and the choice $\Gamma = \ketbra{\Phi}{\Phi}$ maximizes the average entropy.
\end{theorem}

The proof of Theorem~\ref{thm:bounds} is given in Appendix~\ref{app:proof_entropy}. To establish the lower bound, we form a classical--quantum state of smoothed states corresponding to different future measurement outcomes, whose conditional entropy equals the average entropy. We then introduce a specific CPTP map that transforms the smoothed states of the trivial extension into those of any other extension. The data-processing inequality for quantum mutual information then implies the lower bound. The upper bound follows from the concavity of the von Neumann entropy.

\subsection{No universal quantifier for average entropy}
The results above establish that the choice of prior extension
$\Gamma$ strongly affects the average entropy of smoothed state.
It is natural to ask whether there exists a measurement-independent quantifier $Q(\Gamma)$ such that the average entropy is monotone in $Q(\Gamma)$ for
\emph{all} possible measurements. 

We now argue that no such universal quantifier exists due to the incompatibility across POVMs.
Fix a system state $\gamma$ and consider two different extensions
$\Gamma_1,\Gamma_2$ of $\gamma$. While for one POVM $\{F_i\}_i$ it may hold that
$\sum_i p_{F_i}\,S(\rho_{F_i}^{\Gamma_1}) < \sum_i p_{F_i}\,S(\rho_{F_i}^{\Gamma_2})$, where the definitions of $p_{F_i}$ and $\rho_{F_i}^\Gamma$ follow from Eq.~\eqref{eq:p-i} and Eq.~\eqref{eq:rho-i-gamma} respectively, for another POVM $\{G_j\}_j$ the inequality may be reversed to be $\sum_j p_{G_j}\,S(\rho_{G_j}^{\Gamma_1}) > \sum_j p_{G_j}\,S(\rho_{G_j}^{\Gamma_2})$. 

We illustrate this point with a qubit example. Consider two extensions of the maximally mixed state, $\Gamma_1=1/2\,(\ketbra{00}{00}+\ketbra{11}{11})$ and $\Gamma_2=1/2\,(\ketbra{+0}{+0}+\ketbra{-1}{-1})$, where $\ket{0}$ and $\ket{1}$ are eigenstates of Pauli-Z matrix, and $\ket{\pm}=1/\sqrt{2}\,(\ket{0}\pm\ket{1})$ is the eigenstate of Pauli-X matrix. If one performs a Pauli-Z measurement, the average entropy of smoothed state for $\Gamma_1$ is expected to be $0$, while for $\Gamma_2$ it is expected to be $\log 2$. In contrast, if one performs a Pauli-X measurement, the values are reversed: $\log 2$ for $\Gamma_1$ and $0$ for $\Gamma_2$.

This reflects the fact that the average entropy comparison depends not only on the correlation information encoded in $\Gamma$, but also on which ``direction'' in operator space the POVM probes. Thus, any quantifier of extended prior $Q(\Gamma)$ independent of the future measurement cannot determine the orderings of average entropy of smoothed state.

\section{Implications and Connections}\label{sec:implication&connection}
\subsection{Relation to parameter estimation}
The reason for tracing out the auxiliary system $A$ in \cref{eq:smoothig} is simply that we assume people care about nothing but the state of system $Q$. This procedure is indeed not necessary if we would like to preserve any other updated information. Similarly, we call the {\em smoothed global state}
\begin{equation}\label{eq:smoothing_global}
\varrho\sm(t)=\frac{\sqrt{\varrho\fil(t)}\left(\hat{E}\rfil(t)\otimes\one_A\right)\sqrt{\varrho\fil(t)}}{\Tr\left[\rho\fil(t)\hat{E}\rfil(t)\right]}.
\end{equation}
One can therefore estimate the state of any other correlated systems by tracing out the rest. 

As a simple example, here we will show that in the scenario where Alice wants to estimate Bob's outcome till time $t$, this method gives the same result as classical smoothing theory.
We first write down a filtered global state at time $t$ (any extra assumption on the initial state does not alter the result here, so we pick the simplest)
\begin{align}
    \varrho\fil(t)=\sum_{\past{\bf U}_t}\frac{{\Phi}_{\past{\bf O}_t,\past{\bf U}_t}[\rho_0]}{\Tr\left[\Phi_{\past{\bf O}_t}[\rho_0]\right]}\otimes\ketbra{\past{\bf U}_t}{\past{\bf U}_t}.
\end{align}
Plugging this and the definitions of filtered state and retrofiltered effect into \cref{eq:smoothing_global}, we have the corresponding smoothed global state,
\begin{equation}
\sum_{\past{\bf U}_t}\frac{\sqrt{{\Phi}_{\past{\bf O}_t,\past{\bf U}_t}(\rho_0)}\Phi_{\fut{\bf O}_t}^\dagger \left(\one_Q\right)\sqrt{{\Phi}_{\past{\bf O}_t,\past{\bf U}_t}(\rho_0)}}{\Tr\left[\Phi_{\past{\bf O}_t}(\rho_0)\Phi_{\fut{\bf O}_t}^\dagger \left(\one_Q\right)\right]}\otimes\ketbra{\past{\bf U}_t}{\past{\bf U}_t}.
\end{equation}
Tracing out the system $Q$ gives the state of the classical register
\begin{equation}\label{eq:pUt|O}
\begin{split}
    &\Tr_Q(\varrho\sm(t))=\sum_{\past{\bf U}_t}\frac{\Tr\left[{\Phi}_{\past{\bf O}_t,\past{\bf U}_t}(\rho_0)\Phi_{\fut{\bf O}_t}^\dagger \left(\one_Q\right)\right]}{\Tr\left[\Phi_{\past{\bf O}_t}(\rho_0)\Phi_{\fut{\bf O}_t}^\dagger \left(\one_Q\right)\right]}\ketbra{\past{\bf U}_t}{\past{\bf U}_t}\\=&\sum_{\past{\bf U}_t}\frac{p(\past{\bf U}_t,\both{\bf O})}{p(\both{\bf O})}\ketbra{\past{\bf U}_t}{\past{\bf U}_t}=\sum_{\past{\bf U}_t}p(\past{\bf U}_t|\both{\bf O})\ketbra{\past{\bf U}_t}{\past{\bf U}_t}.
\end{split}
\end{equation}
The weight of each set of possible Bob's measurement record is in accordance with classical Bayesian inference.

\subsection{Relation to counterfactual reasoning}

Counterfactual reasoning—asking what \emph{would} have happened had a different choice been made—has long been of fundamental interest in philosophy and science. 
Quantum state smoothing, or retrodiction in general, grounds any counterfactual statement regarding a quantum system within orthodox quantum theory. Our definition of $\rho\sm(t)$ as the optimal estimate \cite{Ge2025MCP,kuang2025quantummeasurementretrodictionentropic} of the system's state at time $t$, given all evidence obtained, provides a natural instantiation of this idea: for an unperformed counterfactual measurement $\{E_j\}_j$, the probability of obtaining outcome $i$ at time $t$ should be calculated according to the Born rule applied to the smoothed state: 
\begin{align}\label{eq:sm_counterfactual}
    p(i) = \Tr[\rho\sm (t) E_i].
\end{align}

The idea presented above is closely related to the recent calculus for quantum counterfactuals introduced by Banerjee et al.~\cite{Banerjee2026counterfactual}. Their work rigorously extends Lewis's analysis~\cite{Lewis1979counterfactual} to the quantum domain by replacing deterministic counterfactual propositions with counterfactual probabilities, which they call \emph{supposabilities}. The calculus is based on the estimation of \emph{fixtures}, all classical events before the counterfactual antecedent, conditioned on evidence obtained in the actual world. To answer the question ``what would Alice measure had she performed a measurement $\{E_j\}_j$ at time $t$, given her observation $\both{\bf O}$'', their formalism marginalizes over the probability distribution of the unknown fixtures:
\begin{equation}\label{eq:blw_supposability}
    \begin{split}
        \text{Su}(i|\{E_j\}_j||\both{\bf O}) =& \sum_{f}p(i|\{E_j\}_j, f)p(f|\both{\bf O})\\
=&\Tr\left[E_i\sum_f p(f|\both{\bf O})\rho(t|f)\right].
    \end{split}
\end{equation}
Here, $\text{Su}$ denotes the supposability. The single vertical bar $(|)$ indicates conditioning on the counterfactual antecedent, and the double vertical bar $(||)$ indicates conditioning on the evidence obtained in the actual world.

When the classical events $f$ determine an ontic quantum state, such as in the Guevara-Wiseman or CLHS scenarios, the averaged state $\sum_f p(f|\both{\bf O})\,\rho(t|f)$ coincides exactly with the corresponding smoothed state. 
In this case, \cref{eq:blw_supposability} reduces to \cref{eq:sm_counterfactual}, showing that the counterfactual probability is precisely the Born probability computed from $\rho\sm(t)$. More generally, our unified retrodictive framework extends this beyond such ontic cases, providing a principled way to assign counterfactual probabilities even when the system’s past cannot be described by an ontic state given all classical information.

This identification highlights a dual role for quantum state smoothing: not only as a tool for inference about the past, but also as the principled foundation for counterfactual predictions.

\section{Conclusion}\label{sec:conclusion}
We have developed a unified framework for quantum state smoothing based on extended retrodiction. By elevating prior beliefs to joint states of the system and auxiliary references, our construction encompasses both Petz-Fuchs and Guevara-Wiseman smoothing as special cases. This resolves the long-standing divide between trajectory-based and retrodictive approaches, and shows that quantum state smoothing is not a few ad hoc constructions but instead arises naturally from quantum Bayesian retrodiction.

Our results clarify that the apparent ambiguity in defining smoothed quantum states originates from different choices of extended priors, which capture different correlations absent from the reduced density matrix description. Our formalism naturally generalizes to broader scenarios where correlations or side information are available, suggesting a rich class of smoothing theories tied to operational assumptions about the prior. In particular, we proved average entropy bounds that are saturated by the Petz-Fuchs and CLHS smoothed states, establishing their extremal status. Furthermore, our framework directly connects smoothing to parameter estimation and to counterfactual reasoning, thus offering immediate practical use.

This work therefore places quantum state smoothing on a firm conceptual footing, aligning it with the logic of Bayesian inference and opening new perspectives for both foundational and applied quantum information. Open questions remain, including the efficient numerical methods for high-dimensional reference systems, and potential applications in quantum error correction, quantum control, and quantum machine learning. We expect that the unified smoothing theory will serve as a versatile tool for these future developments.

\begin{acknowledgments}
M.L. thanks Kiarn T. Laverick for discussions during a previous collaboration that helped inspire parts of this study. This project is supported by the National Research Foundation, Singapore through the National Quantum Office, hosted in A*STAR, under its Centre for Quantum Technologies Funding Initiative (S24Q2d0009). V.S.~acknowledges support from the Ministry of Education, Singapore, under the Tier 2 grant ``Bayesian approach to irreversibility'' (Grant No.~MOE-T2EP50123-0002). G.\ B.\ acknowledges support from the Start-up Fund (Grant No.~G0101000274) from The Hong Kong University of Science and Technology (Guangzhou).
\end{acknowledgments}

\appendix

\section{Proof that the generalized smoothed state satisfies the desiderata}\label{app:proof_smooth}

In this appendix we prove that the proposed {\em generalized quantum smoothed state},
\begin{equation}\label{appeq:smoothing}
\rho\sm(t) = \frac{\Tr_A\!\left[\sqrt{\varrho\fil(t)}\,(\hat{E}\rfil(t)\otimes \one_A)\,\sqrt{\varrho\fil(t)}\right]}{\Tr[\rho\fil(t)\,\hat{E}\rfil(t)]},
\end{equation}
satisfies the three desiderata given in the main text.

\bigskip

{\em Physicality.---}It is positive semidefinite because $\sqrt{\varrho\fil(t)}\,(\hat{E}\rfil(t)\otimes \one_A)\,\sqrt{\varrho\fil(t)}$ can be written as $X^\dag X$, where $X=(\sqrt{\hat{E}\rfil(t)}\otimes \one_A)\,\sqrt{\varrho\fil(t)}$ and the partial trace does not alter positivity. Normalization is guaranteed by the denominator:
\begin{equation}
\begin{split}
\Tr[\rho\sm(t)] &= \frac{\Tr\left[\Tr_A\left[\sqrt{\varrho\fil(t)}(\hat{E}\rfil(t)\otimes \one_A)\sqrt{\varrho\fil(t)}\right]\right]}{\Tr[\rho\fil(t)\hat{E}\rfil(t)]} \\
&= \frac{\Tr\left[\Tr_A[\varrho\fil(t)]\hat{E}\rfil(t)\right]}{\Tr[\rho\fil(t)\hat{E}\rfil(t)]} = 1.
\end{split}
\end{equation}
Thus $\rho\sm$ is Hermitian, positive semidefinite, and unit-trace: a valid quantum state.

\emph{Reduction to Filtering.---}If we average $\rho\sm$ over all possible future records $\fut{\bf O}_t$, conditioned on a fixed past record $\past{\bf O}_t$, we obtain

\begin{equation}
\begin{split}
    &\mathbb{E}_{\fut{\bf O}_t}[\rho_{\text S|\both{\bf O}}(t)|\past{\bf O}_t] = \sum_{\fut{\bf O}_t} p(\fut{\bf O}_t|\past{\bf O}_t) \rho_{\text S|\both{\bf O}}(t)\\=&\sum_{\fut{\bf O}_t} \Tr_A\!\left[\sqrt{\varrho\fil(t)}\,(\Phi_{\fut{\bf O}_t}^\dagger \left(\one_Q\right)\otimes \one_A)\,\sqrt{\varrho\fil(t)}\right]=\rho\fil(t).
\end{split}
\end{equation}

\emph{Classical Limit.---}The limit implicitly states that the filtered global state over system $Q$ and the auxiliary system $A$ is always separable (no quantum correlation). Consider the system dynamics and its measurement operator are diagonal in a fixed orthonormal basis $\ket{x}$. The auxiliary system $A$ is diagonalized in the other orthonormal basis $\ket{a}$. For any specific moment, letting $\wp\fil(x,a)$ and $E\rfil(x)$ denote the corresponding probability vectors, before tracing out auxiliary system $A$,  \cref{appeq:smoothing} reduces to
\begin{equation}
    \varrho\sm= \sum_{x,a}\frac{E\rfil(x)\wp\fil(x,a)}{\sum_{x'} E\rfil(x')\wp\fil(x')} \ketbra{xa}{xa} 
\end{equation}
Ignoring the auxiliary system, it is evident that
\begin{align}
    \rho\sm = \Tr_A [\varrho\sm]=\sum_x\frac{\wp\fil(x)E\rfil(x)}{p(\fut{\bf O}_t|\past{\bf O}_t)}\ketbra{x}{x}.
\end{align}
The probability of the quantum system being at a distinguishable state $\ket{x}$ is exactly the classical smoothed distribution $\wp\sm(x)$. Therefore, the quantum definition reproduces the classical smoothing theory in the appropriate limit.

\bigskip
Hence the generalized smoothed state $\rho\sm$ satisfies all three desiderata.

\section{Proof of bounds on average entropy of generalized smoothed state}\label{app:proof_entropy}
We now prove \cref{thm:bounds}. The argument proceeds in two parts: we first show that the trivial extension minimizes the average entropy, and then that the purification maximizes it.

We first define a linear map $\Lambda:\mathcal S(\Hil_Q)\to\mathcal S(\Hil_Q)$ by
\begin{equation}\label{eq:Lambda-def}
\Lambda(Y):=\Tr_A\left[\sqrt{\Gamma}\,\left(\gamma^{-1/2}\,Y\,\gamma^{-1/2}\otimes \one_A \right)\,\sqrt{\Gamma}\right],
\end{equation}
where $\gamma^{-1/2}$ denotes the inverse on the support of $\gamma$ (if $\gamma$ is full-rank this is the ordinary inverse). The map $\Lambda$ is completely positive because it is a composition of CP maps, and it is trace-preserving on $\supp(\gamma)$ since for any $Y\in\mathcal S(\Hil_Q)$ that has support contained in $\supp(\gamma)$
\begin{equation}
    \begin{split}
        \Tr\Lambda(Y) &= \Tr\left[\sqrt{\Gamma}\,\left(\gamma^{-1/2}Y\gamma^{-1/2}\otimes\one_A\right)\,\sqrt{\Gamma}\right]\\&= \Tr\left[\gamma^{-1/2}Y\gamma^{-1/2}\,\Tr_A\Gamma\right]= \Tr\left(Y\right).
    \end{split}
\end{equation}
Hence $\Lambda$ is a CPTP map on $\supp(\gamma)$.

Next observe that for the particular inputs $\rho_{E_i}^\gamma$, which always have support within $\supp(\gamma)$, we have (by direct substitution)
\[
\Lambda(\rho_{E_i}^\gamma)
= \frac{\Tr_A\!\left[\sqrt{\Gamma}\,(E_i\otimes\one_A)\,\sqrt{\Gamma}\right]}{p_{E_i}} = \rho_{E_i}^\Gamma,\quad\text{for every }i.
\]

Now form the classical--quantum states in system $CQ$
\[
\omega_{CQ}^\gamma=\sum_i p_{E_i}\,\ketbra{i}{i}_C\otimes\rho_{E_i}^\gamma,\quad
\omega_{CQ}^\Gamma=\sum_i p_{E_i}\,\ketbra{i}{i}_C\otimes\rho_{E_i}^\Gamma .
\]
By the intertwining relation above,
\[
\omega_{CQ}^\Gamma=(\mathrm{id}_C\otimes\Lambda)\left(\omega_{CQ}^\gamma\right).
\]
The quantity of interest, the $p$-weighted average von Neumann entropy, equals the conditional entropy:
\begin{align}
    \sum_i p_{E_i}\,S(\rho_{E_i}^\gamma)=S(Q|C)_{\omega_{CQ}^\gamma},\label{appeq:aeeqce1}\\
\sum_i p_{E_i}\,S(\rho_{E_i}^\Gamma)=S(Q|C)_{\omega_{CQ}^\Gamma}.\label{appeq:aeeqce2}
\end{align}
Since $\Lambda$ is CPTP, the data-processing inequality for quantum mutual information \cite{nielsen2010qcqi} implies
\begin{equation}
    S(Q:C)_{\omega_{CQ}^\gamma}\ge S(Q:C)_{(\mathrm{id}_C\otimes\Lambda)\left(\omega_{CQ}^\gamma\right)}=S(Q:C)_{\omega_{CQ}^\Gamma}.
\end{equation}
Or equivalently,
\begin{equation}
    S(\Tr_C\left[\omega_{CQ}^\gamma\right])-S(Q|C)_{\omega_{CQ}^\gamma}\ge S(\Tr_C\left[\omega_{CQ}^\Gamma\right])-S(Q|C)_{\omega_{CQ}^\Gamma}.
\end{equation}
Noticing that $\Tr_C\left[\omega_{CQ}^\gamma\right]=\Tr_C\left[\omega_{CQ}^\Gamma\right]=\gamma$, we have
\begin{align}
    S(Q|C)_{\omega_{CQ}^\gamma} 
    \le\; S(Q|C)_{\omega_{CQ}^\Gamma}.
\end{align}
Using \cref{appeq:aeeqce1,appeq:aeeqce2}, this is equivalent to
\begin{equation}
    \sum_i p_{E_i}\,S(\rho_{E_i}^\gamma) 
      \le \sum_i p_{E_i}\,S(\rho_{E_i}^\Gamma),
\end{equation}
which proves the lower bound.

For the upper bound, we use concavity of the von Neumann entropy:
\[
\sum_i p_{E_i}\,S(\rho_{E_i}^\Gamma)\ \le\ S\!\left(\sum_i p_{E_i}\,\rho_{E_i}^\Gamma\right).
\]
And noticing that
\begin{equation}
    \begin{split}
&\sum_i p_{E_i}\,\rho_{E_i}^\Gamma
= \sum_i \Tr_A\!\big[\sqrt{\Gamma}\,(E_i\otimes\mathbb I_A)\,\sqrt{\Gamma}\big]
\\=& \Tr_A\!\big[\sqrt{\Gamma}\,(\sum_i E_i\otimes\mathbb I_A)\,\sqrt{\Gamma}\big]
= \Tr_A[\Gamma]=\gamma,
    \end{split}
\end{equation}
we have $\sum_i p_{E_i} S(\rho_{E_i}^\Gamma)\le S(\gamma)$.

If $\Gamma$ is pure, write $\Gamma=\ketbra{\Psi}{\Psi}_{QA}$. Since $\sqrt{\ketbra{\Psi}{\Psi}}=\ketbra{\Psi}{\Psi}$ and $\bra{\Psi} (E_i\otimes\one_A)\ket{\Psi} = \Tr[(E_i\otimes\one_A)\Gamma]= p_{E_i}$, we have
\begin{equation}
    \rho_{E_i}^{\ketbra{\Psi}{\Psi}}=\frac{\Tr_A\left[\sqrt{\Gamma}\,(E_i\otimes\one_A)\,\sqrt{\Gamma}\right]}{p_{E_i}}=\Tr_A\left[\ketbra{\Psi}{\Psi}\right] = \gamma.
\end{equation}
It then follows that 
\begin{equation}
    \sum_i p_{E_i} S(\rho_{E_i}^{\ketbra{\Psi}{\Psi}})=\sum_i p_{E_i} S(\gamma)=S(\gamma), 
\end{equation}
showing that $S(\gamma)$ is the maximum and achieved when $\Gamma$ is a purification of $\gamma$. This completes the proof.

\bibliography{refs}

\section*{Data Availability}
Data sharing is not applicable to this article as no datasets were generated or analysed during the current study.

\section*{Author Contributions}
M.L. developed the main framework and wrote the initial draft. G.B. and V.S. supervised the project and contributed to the revision of the manuscript. All authors discussed the results and approved the final manuscript.

\section*{Competing Interests}
The authors declare no competing interests.
\end{document}